# Exploring a quantum-information-relevant magnonic material: ultralow damping at low temperature in the organic ferrimagnet V[TCNE]x

H. Yusuf*[1], M. Chilcote*[1,2], D. R. Candido[3], S. Kurfman[1], D. S. Cormode[1], Y. Lu[1], M. E. Flatté[3], E. Johnston-Halperin[1]

[1]Department of Physics, The Ohio State University, Columbus, Ohio 43210

[2]School of Applied and Engineering Physics, Cornell University, Ithaca, New York 14853

[3]Department of Physics and Astronomy, University of Iowa, Iowa City, Iowa, 52242

* These authors contributed equally to this work.

**Abstract:** Quantum information science and engineering requires novel low-loss magnetic materials for magnon-based quantum-coherent operations. The search for low-loss magnetic materials, traditionally driven by applications in microwave electronics near room-temperature, has gained additional constraints from the need to operate at cryogenic temperatures for many applications in quantum information science and technology. Whereas yttrium iron garnet (YIG) has been the material of choice for decades, the emergence of molecule-based materials with robust magnetism and ultra-low damping has opened new avenues for exploration. Specifically, thin-films of vanadium tetracyanoethylene (V[TCNE]x) can be patterned into the multiple, connected structures needed for hybrid quantum elements and have shown room-temperature Gilbert damping ($\alpha = 4 \times 10^{-5}$) that rivals the intrinsic (bulk) damping otherwise seen only in highly-polished YIG spheres (far more challenging to integrate into arrays). Here, we present a comprehensive and systematic study of the low-temperature magnetization dynamics for V[TCNE]x thin films, with implications for their application in quantum systems. These studies reveal a temperature-driven, strain-dependent magnetic anisotropy that compensates the thin-film shape anisotropy, and the recovery of a magnetic resonance linewidth at 5 K that is comparable to room-temperature values (roughly 2 G at 9.4 GHz).

We can account for these variations of the V[TCNE]$_x$ linewidth within the context of scattering from very dilute paramagnetic impurities, and anticipate additional linewidth narrowing as the temperature is further reduced.

The search for low-loss magnetic materials dates to the early days of radio and microwave electronics [1–3], and the study of elementary excitations, or magnons, in these magnetically-ordered materials has proven to be a rich area of research for both fundamental physics and their potential technological applications. More recently, interest in these low-loss systems has expanded to include applications in the field of quantum information technology such as quantum sensing and quantum transduction [4–7], wherein low-temperature operation allows for the freeze-out of thermal excitations and access to the single-quantum regime. In this regime the field of quantum magnonics utilizes hybrid architectures for coupling magnons to other quantum degrees of freedom, such as microwave photons, with the aim of extending their functionality in the quantum limit [8,9]. It has been demonstrated that magnons can be resonantly excited over a wide range of microwave frequencies, allowing for precise control of qubit states mediated by coherent exchange via cavity-mode photon excitations [4,7]. Magnons also exhibit the potential to coherently couple localized spin-qubits with high cooperativity [10]. However, while magnons exist in a wide range of materials, the same delocalized electrons that are most often responsible for stabilizing ferromagnetic order also contribute to electron-magnon scattering [11], leading to substantial losses in most metallic ferromagnets. As a result, the study of low-dissipation magnon dynamics for quantum applications has focused on insulating ferromagnets and ferrimagnets, with yttrium iron garnet (YIG) and its close relatives holding pride of place as the benchmark low-loss materials for more than 50 years [4,12–14]. As a result, despite these longstanding and emerging needs, applications are still constrained by the materials limitations of YIG; namely the need for growth or annealing at high temperatures (typically 800°C) [15–17] and the resulting difficulty in integrating and patterning YIG thin-films with other microwave electronic structures and devices.

In this context, the emergence of the molecule-based ferrimagnet vanadium tetracyanoethylene (V[TCNE]$_x$) has dramatically expanded the playing field for low-loss magnets. Despite what one might expect from its molecular building blocks, V[TCNE]$_x$ has a magnetic ordering temperature of over 600 K and shows sharp hysteresis at room-temperature [18–20]. Moreover, its dynamic properties are exceptional, showing ultra-narrow ferromagnetic resonance (FMR) linewidth (typically ~ 1 – 1.5 G at 9.4 GHz) with a Gilbert damping parameter, $\alpha$, of $4 \times 10^{-5}$ for thin-films [18,21]. As a comparison, the best YIG thin-films typically show $\alpha = 6.5 \times 10^{-5}$ [22] and a value of $4 \times 10^{-5}$ is competitive with the *intrinsic* damping of bulk YIG $\alpha = 3 \times 10^{-5}$ [15,23]. From an applications perspective, V[TCNE]$_x$ has been shown to deposit on a wide variety of substrates without compromising material quality [24–26], facile encapsulation allows for direct integration with pre-patterned microwave structures for operation under ambient conditions [27], and recent work has demonstrated patterning at length scales down to 10 μm without increased damping [21]. However, while these properties clearly establish the potential of V[TCNE]$_x$ for new applications in traditional microwave electronics, very little is known about its low-temperature magnetization dynamics and therefore its potential for applications in quantum information science and engineering (QISE).

Here we present a detailed study of the low-temperature magnetic resonance of V[TCNE]$_x$ films. We identify two regimes. In the high-temperature regime, extending from 300 K down to 9 K, we observe a monotonic shift in the resonance frequency consistent with a temperature-dependent strain. This strain results in a crystal-field anisotropy that increases with decreasing temperature with a magnitude of at least 140 Oe and the same symmetry, but opposite sign, to the shape anisotropy of the thin-film. In addition, we observe an increase in linewidth consistent with magnon scattering from paramagnetic impurities similar to what has been observed in YIG [23,28,29], but with an amplitude 3 times smaller (*i.e.* an increase in linewidth by 9 times in V[TCNE]$_x$ as compared to 28 times in YIG [23,30]). In the low-temperature regime, starting at 9 K and extending to 5 K, we observe a discontinuous change in both anisotropy and linewidth: the anisotropy abruptly reverts to the room-temperature symmetry (in-plane easy-axis) and the linewidth approaches room-temperature values (2.58 G) at 5 K. This linewidth variation can be explained using a model for scattering between magnons and paramagnetic impurities that takes into account the finite spin-lifetime of the impurity spins [23,31]. At high temperatures (above 100 K) the spin-lifetime is sufficiently short that changes in temperature do not lead to significant changes in scattering rate, and at low-temperatures (below 9 K) the spin-lifetime becomes long with respect to the spin-magnon scattering time, resulting in a saturation of the excited state. At intermediate temperatures (from 9 K to 100 K) this spin-magnon scattering dominates relaxation due to the increase of the ground state impurity population, which results in a local maximum in the linewidth that is 9 times larger than the room-temperature value. These results are extremely promising for low-temperature applications of V[TCNE]$_x$ magnonics, promising low-temperature magnon resonators with unprecedented low-loss that can be integrated on-chip into microwave electronic circuits and devices [20,21].

For this study, thin-films of V[TCNE]$_x$ are deposited on sapphire (Al$_2$O$_3$ (0001)) substrates using chemical vapor deposition (CVD) growth process consistent with prior reports [18,19]. Briefly, argon gas transfers the two precursors tetracyanoethylene (TCNE) and vanadium hexacarbonyl (V(CO)$_6$) into the reaction zone of a custom-built CVD reactor (Fig. 1(a)) where V[TCNE]$_x$ is deposited onto polished sapphire substrates. The system is temperature controlled to maintain the TCNE, V(CO)$_6$ and the reaction zones at 65° C, 10° C and 50° C respectively. After growth the sample is mounted on a custom, microwave-compatible sample holder and sealed using a septa cap in an electron paramagnetic resonance (EPR) grade quartz tube in an argon environment. When the sample is not being measured, it is stored in a -35° C freezer housed in an argon glovebox and is stable for over one month [27].

Ferromagnetic resonance (FMR) measurements are performed using a Bruker EMX Plus X-band EPR spectrometer at temperatures ranging from 300 K down to 5 K. The microwave frequency of the spectrometer is tuned between 9 and 10 GHz for optimal microwave cavity performance before the measurement, and then the frequency is fixed while the DC field is swept during data collection. Figure 1(b) shows a representative room-temperature FMR measurement of a typical V[TCNE]$_x$ thin-film with the external magnetic field applied in the plane of the sample. The resonance feature is consistent with previously reported high-quality V[TCNE]$_x$ thin-film growth, showing a peak-to-peak linewidth of 1.5 G at 9.4 GHz [18,19].

Comparing this data to FMR measurements at temperatures of 80 K and 40 K (Figure 1(c)) shows an increase in the resonance field of over 40 G (roughly half of the saturation magnetization, $4\pi M_s$) as the temperature decreases. Since the applied microwave frequency is held constant at 9.4 GHz, this shift must arise from fields internal to the V[TCNE]$_x$ film, i.e. magnetic anisotropy fields. Note that since the value of the DC applied field varies between 3350 G and 3450 G, well above $4\pi M_s$, changes in the magnetization of the film are not expected to contribute to this field shift. In a similar fashion, changes in the shape-dependent anisotropy fields can be ruled out, leaving only changes to the crystal-field anisotropy as a potential source of this phenomenon. Crystal-field anisotropy originates from the interaction of a material's mean exchange field and the angular momenta of neighboring atoms (ions) in the material, indicating that there is a temperature dependence to the local atomic environment within the V[TCNE]$_x$ films, *e.g.* due to a temperature-dependent strain within the film.

In order to more comprehensively map out this phenomenon angle dependent FMR measurements are performed to quantitatively track changes in the magnetic anisotropy at temperatures of 300 K, 80 K, and 40 K (Fig. 2). Variation of the magnetic resonance field as a function of the angle between the applied field and the principal axes of the film can be modeled by considering the free energy of the magnetic system with anisotropic contributions. If we consider the case of a uniaxial anisotropy with the hard-axis perpendicular to the easy-axis, and where the magnetization is parallel to the external field (*i.e.* external field is much larger than the saturation magnetization, as is the case here) the total magnetostatic energy is as follows [32]:

$$E = -\boldsymbol{M} \cdot \boldsymbol{H} + 2\pi(\boldsymbol{M} \cdot \boldsymbol{n})^2 - K(\boldsymbol{M} \cdot \boldsymbol{u}/M)^2 \qquad (1)$$

where $\boldsymbol{M}$ is the magnetization, $\boldsymbol{H}$ is the applied magnetic field, $\boldsymbol{n}$ is the unit vector parallel to the normal of the magnetic sample, $\mathbf{u}$ is the unit vector parallel to the easy-axis and $K$ is an anisotropy constant. For the case of in-plane uniaxial anisotropy, this simplifies to

$$E = -MH\left(\sin\phi\sin^2\theta + \cos^2\theta\right) + 2\pi M^2\cos^2\theta - K\sin^2\theta\sin\phi^2 \qquad (2)$$

where $\theta$ is the angle between $\boldsymbol{M}$ and the sample normal and $\phi$ is the azimuthal angle. Minimizing the magnetostatic energy with respect to $\theta$ , one will find that the easy-axis orientation occurs when $\theta = 2n\pi \pm \frac{\pi}{2}$, where $n$ is an integer. Using this simple symmetry analysis, we can see that the data in Fig. 2 indicates that the easy-axis lies in-plane at a temperature of 300 K (i.e. the resonance field is smallest when the applied magnetic field lies in-plane) and out-of-plane at a temperature of 40 K (i.e. the resonance field is smallest when the applied magnetic field is out-of-plane). In this context, the lack of variation in resonance field at 80 K indicates a nearly isotropic magnetic response. This switch in magnetic easy-axis from in-plane to out-of-plane further supports the proposition that there is an additional temperature-dependent crystal-field contribution to the magnetic anisotropy.

In previous studies, templated growth of V[TCNE]$_x$ resulting in nanowire morphologies induced an additional in-plane magnetic anisotropy with easy-axis perpendicular to the long-axis of the nanowires, strongly suggesting the presence of a strain-dependent contribution to the crystal-field anisotropy [33]. In the thin-films studied here, such a strain-dependent crystal-field effect would be expected to generate anisotropy parallel to the surface normal, *i.e.* in the out-of-plane direction. The anisotropy field would then be parallel to the expected shape anisotropy from a thin-film, though not necessarily with the same sign. As a result, if there is a difference in the coefficient of thermal expansion between the V[TCNE]$_x$ film and the sapphire substrate then the temperature dependence of magnetic anisotropy can potentially be understood as a proxy for a temperature dependence of strain in the thin-film; such variations in strain leads to changes in the local atomic structure, leading to the observed changes in magnetic anisotropy. We note that while the coefficient of thermal expansion for V[TCNE]$_x$ has not yet been measured, the value for sapphire is 5.4 ppm/K and typical values for molecular-based solids can range somewhere between 28−500 ppm/K [34]. Assuming no strain at room-temperature, this would then imply a compressive strain between 0.67% to 15% at the sapphire−V[TCNE]$_x$ interface at 5 K, leading to an out-of-plane distortion whose symmetry is consistent with the observed anisotropy.

A schematic describing how these two anisotropy fields would be expected to interact as a function of temperature can be found in Fig. 3(a). At a temperature of 300 K (Fig. 3(a), upper panels), the orientation of the easy-axis is determined by the shape anisotropy, resulting in an in-plane easy-axis for thin-films. But at a temperature of 40 K (Fig. 3(a) lower panels), there is an additional crystal-field anisotropy, $H_\perp$, proposed that dominates the shape anisotropy, reorienting the easy-axis to be out-of-plane. This symmetry analysis also explains the lack of orientation dependence at a temperature of 80 K, which is apparently the temperature at which the strain-driven crystal-field anisotropy perfectly cancels out the shape anisotropy. We note that similar phenomenology is also observed in vanadium methyl tricyanoethylenecarboxylate (V[MeTCEC]$_x$) thin-films (see supplementary materials), indicating that this temperature- and strain-dependent anisotropy is a general property of this class of metal-ligand ferrimagnets.

The fact that the shape and proposed crystal-field anisotropies have the same symmetry make it challenging to distinguish between the two; therefore, an effective field is defined as $H_{eff} = 4\pi M_{eff} = 4\pi M_s - H_\perp$, where $M_s$ is the saturation magnetization and $H_\perp$ is the crystal-field anisotropy. Figure 2 shows the effects of this net anisotropy field in the form of resonance field shifts and a change in the easy-axis orientation. Quantitatively extracting the magnitude and direction of this anisotropy field provides detailed insight into the role of crystal-field anisotropy in tuning the magnetic response of V[TCNE]$_x$ thin-films. To this end, each scan is fit to the sum of the derivatives of absorption and dispersion from a Lorentzian function to extract the resonance frequency and linewidth (experimental data are obtained using a modulated-field technique that yields the derivative of the expected Lorentzian resonance lineshape). For scans showing an out-of-plane easy-axis a single derivative sum provides good agreement with the data, while for scans showing in-plane easy-axis more complex structure is observed requiring the addition of up to three derivative sums. In the results discussed below we focus on the behavior of the primary, *i.e.* largest amplitude, peak (a full description of the fitting and resulting phenomenology can be found in the supplemental material).

Figure 3(b) shows the extracted resonance field plotted against sample rotation angle for the high-and low-temperature data shown in Fig. 2, 300 K and 40 K, respectively. Taking into account a uniaxial out-of-plane anisotropy defined by $H_{eff}$, as described above, the angular dependence for in-plane to out-of-plane rotation of a thin-film sample is given by [19,35,36]:

$$\frac{\omega}{\gamma} = \sqrt{(H - H_{eff} \cos^2 \theta)(H - H_{eff} \cos 2\theta)}$$
$$= \sqrt{(H - (4\pi M_s - H_\perp) \cos^2 \theta)(H - (4\pi M_s - H_\perp) \cos 2\theta)} \quad (3)$$

where $\omega$ is the resonance frequency and $\gamma$ is the gyromagnetic ratio. As a result, the phenomenology of the data presented in Fig. 2 can be understood as an $H_{eff}$ that is positive at 300 K and negative at 40 K, as $H_\perp$ increases with decreasing temperature, consistent with the mechanism for anisotropy switching described in Fig. 3(a). This qualitative understanding can be made quantitative by fitting the data in Fig. 2 using Eq. (3) to extract $H_{eff} = 4\pi M_{eff}$ of 91.2 G $\pm$ 1.6 G and -22.8 G $\pm$ 0.4 G, respectively.

Figure 3(c) shows this $H_{eff}$ plotted against temperature over the temperature range from 300 K to 5 K, extracted from angular dependencies such as the measurements presented in Fig. 2. It should be noted that each anisotropy point in Figure 3(c) represents a fit to a complete angular dependence such as the data shown in Figure 3(b).The effective field makes a smooth transition through zero from positive (in-plane) to negative (out-of-plane) at a temperature of roughly 80 K. This behavior is qualitatively consistent with the phenomenological model presented above and reveals a magnitude of the variation in $H_{eff}$, from +91.2 G ± 1.6 G at 300 K to -45.2 G ± 1.1 G at 10 K, that is roughly 150% of the room-temperature value.

Notably, this more comprehensive study also reveals new phenomenology at the lowest temperature of 5 K, where the anisotropy abruptly shifts back to in-plane with a value of +26.2 G ± 0.6 G (roughly 25% of the room-temperature value). This behavior reproduces across all samples measured and is quantitatively reproduced upon temperature cycling of individual films. The abruptness of this change is distinct from the broad and monotonic behavior observed for temperatures greater than 9 K. The origin of this abrupt change is unclear, but there are two potential explanations consistent with this phenomenology. First, it is possible that the increase in strain results in an abrupt relaxation through the creation of structural defects. This explanation would require some level of self-healing upon warming in order to explain the reproducibility of the transition. Given the lack of long-range structural order in V[TCNE]$_x$ films as-grown [37] it is possible that any residual structural defects do not contribute to additional magnetic loss (damping). Second, it is possible that there exist paramagnetic spins in the system that magnetically order at temperatures below 9 K. If such spins were preferentially located in an interface layer, for example, their ordering could create an exchange bias that would then pull the easy-axis back to an in-plane orientation.

The temperature dependence of the linewidth of the magnetic resonance provides an additional avenue for evaluating these potential explanations. Figure 4 shows the linewidth for the in-plane magnetic resonance from 300 K to 5 K, with additional data to more clearly resolve the sharp change between 5 K and 9 K. The linewidth data presented in Figure 4 is extracted from a single (in-plane applied magnetic field orientation) scan. As a result, the initial dataset underlying Figure 3 was supplemented by a second temperature dependent scan at fixed angle in Figure 4. This data reveals a monotonic increase in linewidth with decreasing temperature from 300 K down to 9 K followed by a dramatic decrease in linewidth between 9 K and 5 K, coincident with the abrupt change in magnetic anisotropy. We note that in studies of YIG thin-films broadly similar phenomenology is observed, though with a maximum in linewidth that is both higher amplitude (roughly 28 times the room-temperature value) and at higher temperature (typically 25 K) than is observed here [23,30]. Prior work [23,28] has explained this behavior using a model of magnon scattering from paramagnetic defect spins (also referred to as two-level fluctuators, TLF) wherein the scattering cross-section at high temperature increases with decreasing temperature as the thermal polarization of the spins increases. This phenomenology competes with magnon-pumping of the paramagnetic spins into their excited state, a process that saturates as the spin-lifetime of the defects becomes long relative to the spin-magnon scattering time. The competition between these two processes yields a local maximum in the damping (linewidth) that depends on the temperature dependent spin lifetime, $t_s$, the energy separation between majority and minority spin states, $\hbar\omega_{eg}$, and the difference between that energy splitting and the uniform magnon energy, $(\hbar\omega - \hbar\omega_{eg})$.

In this model, the linewidth expression is proportional to the square of the exchange interaction energy between V[TCNE]$_x$ atoms and the impurity level $(\hbar\omega_{int})^2 \sim (\hbar\omega_{eg})^2$, a line-shape factor accounting for the finite spin lifetime, $1/t_s/(\hbar^2/t_s^2 + (\hbar\omega - \hbar\omega_{eg})^2 t_s^2)$, and the ratio between the ground and excited impurity states for fast impurity relaxation, $tanh(\hbar\omega/2k_BT)$ [23, 28],

$$\Delta H = \frac{S}{\gamma} \frac{N_{imp}}{N} (\hbar\omega_{int})^2 \frac{1/t_s}{\hbar^2/t_s^2 + (\hbar\omega - \hbar\omega_{eg})^2 t_s^2} \tanh\left(\frac{1}{2}\frac{\hbar\omega}{k_BT}\right) + H_O \quad (4)$$

where $N_{imp}/N$ is the ratio between number of impurities and number of V[TCNE]$_x$ atoms, and $S$ is the averaged V[TCNE]$_x$ spin per site, $\gamma$ is the gyromagnetic ratio and $H_O$ is a constant offset due to other relaxation mechanisms. In addition, we assume spin lifetime $t_s$ $= t_\infty e^{E_b/k_B T}$ [31, 38, 39] where $t_\infty$ is the spin lifetime limit at very high temperatures, and $E_b$ is a phenomenological activation energy. Figure 4 includes a fit of Eq. (4) to the experimental linewidth (orange line) that yields for S ~ 1 and $\omega_{int} \sim \omega_{eg}$ the parameters: $\omega_{eg} t_\infty = 0.98$, $E_b = 1\text{meV}$, $\omega_{eg} N_{imp}/N = 36.5\text{GHz}$ and $H_o = 1$ G. Interestingly, if we assume a reasonable value for $\hbar\omega_{eg}$ of 1.3 meV, a value of $N_{imp}/N = 0.1$ follows, thus indicating that V[TCNE]$_x$ is an exceptional low-loss magnetic material even if we assume an impurity concentration as high as 10%. This observation is consistent with the hypothesis of insensitivity to structural defects discussed above.

However, it is important to note that the peak in linewidth coincides with the abrupt reversion in anisotropy from an out-of-plane easy-axis to an in-plane easy-axis. This change in magnetic anisotropy has the potential to have a substantial impact on spin-magnon scattering efficiency. For example, this change will result in a shift of the energy of the magnon bands (see Eq. 1), and if this change involves a commensurate change in the strain there will also be a modification to the spin-orbit coupling and exchange parameters at the paramagnetic defects. It should be noted that although this reentrant anisotropy is an intriguing feature, the fits to our model for TLFs in Figure 4 are able to reproduce our linewidth data without reference to this effect. As a result, we interpret this fit as an upper bound on $E_b$. This is represented by the additional fits shown in Figure S7 within the Supplemental Material wherein we assume a lower temperature for the nominal peak in linewidth occurring due to spin-magnon scattering that is experimentally preempted by the change in magnetic anisotropy. These alternate fits agree with experimental observations at temperatures above 9 K, and therefore must be considered as possible mechanisms. Moreover, if the residual paramagnetic spins are ordered at temperatures below 9 K, one would require a large amount of energy (>> $\hbar\omega$) to populate their excited states, which is unlikely to happen. Hence, magnetic ordering of the paramagnetic spins would also enhance the suppression of spin-magnon scattering, resulting in the sharp linewidth suppression for $T < 9$ K.

When considering the expected behavior as the temperature is further reduced below 5 K, as would be the case for many applications in QISE, it is useful to consider recent milliKelvin-range measurements of YIG films [40]. That work confirms the expected continued narrowing down to 500 mK followed by a modest increase from 500 mK down to 20 mK, for an overall line narrowing of roughly a factor of 2. The model of scattering from TLFs described above is consistent with this result in YIG if one supposes a second population of TLFs that are dipole coupled to the magnons rather than exchange coupled, for example dilute magnetic impurities in the substrate or environment. We note that extending this model into V[TCNE]$_x$ requires taking into account: i) the substantial difference in structure and chemistry between V[TCNE]$_x$ and YIG, and ii) the fact that $M_s$ in V[TCNE]$_x$ is roughly 20 times smaller than in YIG. The former consideration indicates that the presence of these dipole coupled TLFs need not correlate between the two systems, while the latter predicts that any relaxation associated with their presence should be reduced by a factor of 20 from Ref. [40]. As a result, the overall factor of 2 decrease in linewidth observed in YIG between temperatures of 5 K and 20 mK should be taken as an extremely conservative lower bound on the performance of V[TCNE]$_x$. Given that the linewidth in V[TCNE]$_x$ at 5 K is already on par with its room temperature value, these results firmly establish the suitability for this material for applications in quantum magnonics and related aspects of QISE.

In conclusion, this work presents the first systematic study of the magnetization dynamics of V[TCNE]$_x$ at low temperatures. A strong variation in resonance frequency and anisotropy with temperature is observed, and attributed to a temperature-dependent strain arising from the mismatch in thermal expansion coefficients between V[TCNE]$_x$ films and their sapphire substrates. The resonance linewidth of these films is found to increase with decreasing temperature up to a maximum value of 15 G (roughly 9 times the room-temperature value) and is well fit by a model based on magnon scattering from paramagnetic defect spins. At 5 K the magnetic anisotropy reverts to in-plane, coinciding with a nearly complete recovery of the resonance linewidth to room-temperature values; quantitative modeling suggests the linewidth behavior arises from scattering from paramagnetic defect spins that is suppressed at very low-temperature. This suppression of spin-magnon scattering is expected to strengthen as temperature is further decreased into the milli-Kelvin range due to freeze-out of thermal magnons and phonons, providing a compelling case for the utility of V[TCNE]$_x$ for low-temperature microwave applications, such as those emerging in the field of quantum information science and technology.

**Acknowledgements:** The authors would like to thank A. Franson for providing a software suite for fitting FMR spectra as well as general fitting assistance, and G. Fuchs for fruitful discussions. The work presented in the main text, both experiment and theory, was primarily supported by the U.S. Department of Energy, Office of Basic Energy Sciences, under Award Number DE-SC0019250. S. Kurfman was supported by NSF EFMA-1741666 and grew V[TCNE]$_x$ calibration samples used for preliminary measurements not explicitly included in this paper. Work on V[MeTCEC]$_x$ presented in the supplementary material was performed by M. Chilcote and Y. Lu with the support of NSF Grant No. DMR- 1741666.

**Data availability statement:** See supplementary material at URL will be inserted by AIP Publishing for datasets pertaining to temperature-dependent anisotropy of V[MeTCEC]$_x$, method for extracting linewidth of V[TCNE]$_x$ from FMR scans and additional fits to experimental data highlighting temperature dependence of V[TCNE]$_x$ linewidth.

### References:

[1]    A. Raveendran, M. T. Sebastian, and S. Raman, "Applications of Microwave Materials: A Review" J. Electron. Mater. **48**, 2601 (2019).

[2]    Ü. Özgür, Y. Alivov, and H. Morkoç, "Microwave ferrites, part 1: Fundamental properties" J. Mater. Sci. Mater. Electron. **20**, 789 (2009).

[3]  J. M. Silveyra, E. Ferrara, D. L. Huber, and T. C. Monson, "Soft magnetic materials for a sustainable and electrified world" Science. **362**, (2018).

[4]  Y. Tabuchi, S. Ishino, A. Noguchi, T. Ishikawa, R. Yamazaki, K. Usami, and Y. Nakamura, "Coherent coupling between a ferromagnetic magnon and a superconducting qubit" Science **349**, 405-408 (2015).

[5]  E. Lee-Wong, R. Xue, F. Ye, A. Kreisel, T. Van Der Sar, A. Yacoby and C. R. Du, "Nanoscale Detection of Magnon Excitations with Variable Wavevectors Through a Quantum Spin Sensor" Nano Lett. **20** (5), 3284-3290 (2020).

[6]  R. G. E. Morris, A. F. Van Loo, S. Kosen and A. D. Karenowska, "Strong coupling of magnons in a YIG sphere to photons in a planar superconducting resonator in the quantum limit" Sci Rep. Mater. **7** (1), 11511 (2017).

[7]  S. P. Wolski, D. Lachance-Quirion, Y. Tabuchi, S. Kono, A. Noguchi, K. Usami and Y. Nakamura and E. Wahlström, "Dissipation-Based Quantum Sensing of Magnons with a Superconducting Qubit" Phys. Rev. Lett. **125**, 117701 (2020).

[8]  D. Lachance-Quirion, Y. Tabuchi, A. Gloppe, K. Usami and Y. Nakamura, "Hybrid quantum systems based on magnonics" Appl. Phys. Express **12**, 070101 (2019).

[9]  S. Kosen, R. G. E. Morris, A. F. Van Loo and A. D. Karenowska, "Measurement of a magnonic crystal at millikelvin temperatures" Appl. Phys. Lett. **112**, 012402 (2018).

[10]  D. R. Candido, G. D. Fuchs, E. Johnston-Halperin and M. E. Flatté, "Predicted strong coupling of solid-state spins via a single magnon mode" Mater. Quantum.

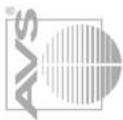
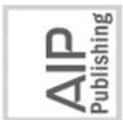

Technol. **1**, 011001 (2021).

[11]  V. S. Lutovinov and M. Y. Reizer, "Relaxation processes in ferromagnetic metals" Zh. Eksp. Teor. Fiz. **77**, 707-716 (1979).

[12]  D. Lachance-Quirion, S. P. Wolski, Y. Tabuchi, S. Kono, K. Usami, and Y. Nakamura, "Entanglement-based single-shot detection of a single magnon with a superconducting qubit" Science. **367**, 425 (2020).

[13]  R. G. E. Morris, A. F. Van Loo, S. Kosen, and A. D. Karenowska, "Strong coupling of magnons in a YIG sphere to photons in a planar superconducting resonator in the quantum limit" Sci. Rep. **7**, (2017).

[14]  M. Kostylev and A. A. Stashkevich, "Proposal for a microwave photon to optical photon converter based on traveling magnons in thin magnetic films" J. Magn. Magn. Mater. **484**, 329 (2019).

[15]  M. C. Onbasli, A. Kehlberger, D. H. Kim, G. Jakob, M. Kläui, A. V. Chumak, B. Hillebrands, and C. A. Ross, "Pulsed laser deposition of epitaxial yttrium iron garnet films with low Gilbert damping and bulk-like magnetization" APL Mater. **2**, (2014).

[16]  S. A. Manuilov and A. M. Grishin, "Pulsed laser deposited Y3Fe5O12 films: Nature of magnetic anisotropy II" J. Appl. Phys. **108**, (2010).

[17]  S. A. Manuilov, R. Fors, S. I. Khartsev, and A. M. Grishin, "Submicron Y3Fe5O12 film magnetostatic wave band pass filters" J. Appl. Phys. **105**, (2009).

[18]  M. Harberts, Y. Lu, H. Yu, A. J. Epstein, and E. Johnston-Halperin, "Chemical Vapor Deposition of an Organic Magnet, Vanadium Tetracyanoethylene" J. Vis.

Exp. (2015).

[19]  H. Yu, M. Harberts, R. Adur, Y. Lu, P. C. Hammel, E. Johnston-Halperin, and A. J. Epstein, "Ultra-narrow ferromagnetic resonance in organic-based thin films grown via low temperature chemical vapor deposition" Appl. Phys. Lett. **105**, 012407 (2014).

[20]  N. Zhu, X. Zhang, I. H. Froning, M. E. Flatté, E. Johnston-Halperin, and H. X. Tang, "Low loss spin wave resonances in organic-based ferrimagnet vanadium tetracyanoethylene thin films" Appl. Phys. Lett. **109**, 082402 (2016).

[21]  A. Franson, N. Zhu, S. Kurfman, M. Chilcote, D. R. Candido, K. S. Buchanan, M. E. Flatté, H. X. Tang, and E. Johnston-Halperin, "Low-damping ferromagnetic resonance in electron-beam patterned, high- Q vanadium tetracyanoethylene magnon cavities" APL Mater. **7**, (2019).

[22]  C. Hauser, T. Richter, N. Homonnay, C. Eisenschmidt, M. Qaid, H. Deniz, D. Hesse, M. Sawicki, S. G. Ebbinghaus, and G. Schmidt, "Yttrium Iron Garnet Thin Films with Very Low Damping Obtained by Recrystallization of Amorphous Material" Sci. Rep. **6**, 1 (2016).

[23]  M. Sparks, *Ferromagnetic-Relaxation Theory* (McGraw Hill, New York, 1964) p. 226

[24]  D. De Caro, M. Basso-Bert, J. Sakah, H. Casellas, J. P. Legros, L. Valade, and P. Cassoux, "CVD-grown thin films of molecule-based magnets" Chem. Mater. **12**, 587 (2000).

[25]  J. M. Manriquez, G. T. Yee, R. S. McLean, A. J. Epstein, and J. S. Miller, "A

Room-Temperature Molecular/Organic-Based Magnet" Science (80-. ). **252**, 1415 LP (1991).

[26] K. I. Pokhodnya, A. J. Epstein, and J. S. Miller, "Thin-Film V[TCNE]x Magnets" Adv. Mater. **12**, 410 (2000).

[27] I. H. Froning, M. Harberts, Y. Lu, H. Yu, A. J. Epstein, and E. Johnston-Halperin, "Thin-film encapsulation of the air-sensitive organic-based ferrimagnet vanadium tetracyanoethylene" Appl. Phys. Lett. **106**, (2015).

[28] P. E. Seiden, "Ferrimagnetic resonance relaxation in rare-earth iron garnets" Phys. Rev. **133**, A728 (1964).

[29] A. M. Clogston, "Relaxation Phenomena in Ferrites" Bell Syst. Tech. J. **34**, 739 (1955).

[30] C. L. Jermain, S. V. Aradhya, N. D. Reynolds, R. A. Buhrman, J. T. Brangham, M. R. Page, P. C. Hammel, F. Y. Yang, and D. C. Ralph, "Increased low-temperature damping in yttrium iron garnet thin films" Phys. Rev. B **95**, 174411 (2017).

[31] W. A. Yager, J. K. Galt, and F. R. Merritt, "Ferromagnetic resonance in two nickel-iron ferrites" Phys. Rev. **99**, 1203 (1955).

[32] H. Puszkarski and M. Kasperski, *On the Interpretation of the Angular Dependence of the Main FMR/SWR Line in Ferromagnetic Thin Films* (2012).

[33] M. Chilcote, M. Harberts, B. Fuhrmann, K. Lehmann, Y. Lu, A. Franson, H. Yu, N. Zhu, H. Tang, G. Schmidt, and E. Johnston-Halperin, "Spin-wave confinement and coupling in organic-based magnetic nanostructures" APL Mater. **7**, (2019).

[34] Y. Mei, P. J. Diemer, M. R. Niazi, R. K. Hallani. K. Jarolimek, C. S. Day, C.

Risko, J. E. Anthony, A. Amassian and O. D. Jurchescu, "Crossover from band-like to thermally activated charge transport in organic transistors due to strain-induced traps" PNAS **114**, 33 (2017).

[35] H. Suhl, "Ferromagnetic Resonance in Nickel Ferrite Between One and Two Kilomegacycles" Phys. Rev. **97**, 555 (1955).

[36] J. Smit and H. G. Beljers., "Ferromagnetic resonance absorption in $BaFe_{12}O_{19}$" Philips Res. Rep. 10, 113 (1955).

[37] M. Chilcote, Y. Lu, and E. Johnston-Halperin, *Organic-Based Magnetically Ordered Films* (World Scientific, 2018).

[38] J. K. Galt and E. G. Spencer, "Loss Mechanism in Spinel Ferrites" Phys. Rev. 127, 1572, 1962.

[39] H. Maier-Flaig, S. Klingler, C. Dubs, O. Surzhenko, R. Gross, M. Weiler, H. Huebl, and S. T. B. Goennenwein, "Temperature dependent damping of yttrium iron garnet spheres" Phys. Rev. B 95, 214423 (2017).

[40] S. Kosen, A. F. van Loo, D. A Bozhko, L. Mihalceanu, R. Gross, and A. D. Karenowska, "Microwave magnon damping in YIG films at millikelvin temperatures" APL Mater. 7, 101120 (2019).

**Figure Legends:**

**Figure 1**

(a) Schematic (planar view) of the CVD growth system; (b) FMR scan of V[TCNE]$_x$ thin film at 300 K with the applied magnetic field applied in the plane (IP) of the sample with $\theta = 90^o$ and resonance frequency of 9.4 GHz. $\Delta$H$_{pp}$ denotes the peak-to-peak linewidth measured as the difference between the positive and negative peak positions; (c) FMR line scans for in-plane field orientation at 300 K, 80 K and 40 K with $\theta = 90^o$ and resonance frequency of 9.4 GHz.

**Figure 2**

Angle-dependent FMR spectra at temperatures of 300 K, 80 K and 40 K at different field orientations with respect to the sample normal. Nominally the sample is rotated from $\theta = -10^o$ to $\theta = 100^o$ in increments of $10^o$, where $\theta = 90^o$ and $\theta = 0^o$ are in-plane and out-of-plane field orientations respectively. Angle corrections have been taken into account (through fitting with Eq. (3)) to reflect the actual rotation angles, denoted by the black arrows to the right of each of the temperature-labeled panels.

**Figure 3**

(a) Schematic of the changes in anisotropy at 300 K and 40 K. $\boldsymbol{H_{app}}$ denotes the external magnetic field, $\boldsymbol{H_{demag}}$ represents the demagnetizing field of the V[TCNE]$_x$ film and $\boldsymbol{H_{crystal}}$ is the crystal-field anisotropy. It should be noted that a finite thin-film has a (negligibly) small demagnetization field when the external filed is applied in the plane since this is not a truly infinite film; (b) Resonance field at different field orientations plotted against sample rotation angles for 300 K (open circles) and 40 K (filled circles) and fits to Eq. (3) (dashed and solid line, respectively) to extract the effective field $H_{eff}$; (c) $H_{eff}$ plotted against temperature ranging from 300K – 5K. The inset shows the FMR lineshapes at 300 K and 5 K; fitting the data to extract the linewidth at FWHM gives 1.63 G and 2.58 G respectively, this shows that the two linewidths are indeed comparable with the linewidth at 5 K only about 1.66 times larger than the room-temperature value. For both (b) and (c), experimental errors are smaller than the point size.

**Figure 4**

V[TCNE]$_x$ linewidth as a function of temperature (black points) and corresponding curve fit (orange line) using Eq. (4).

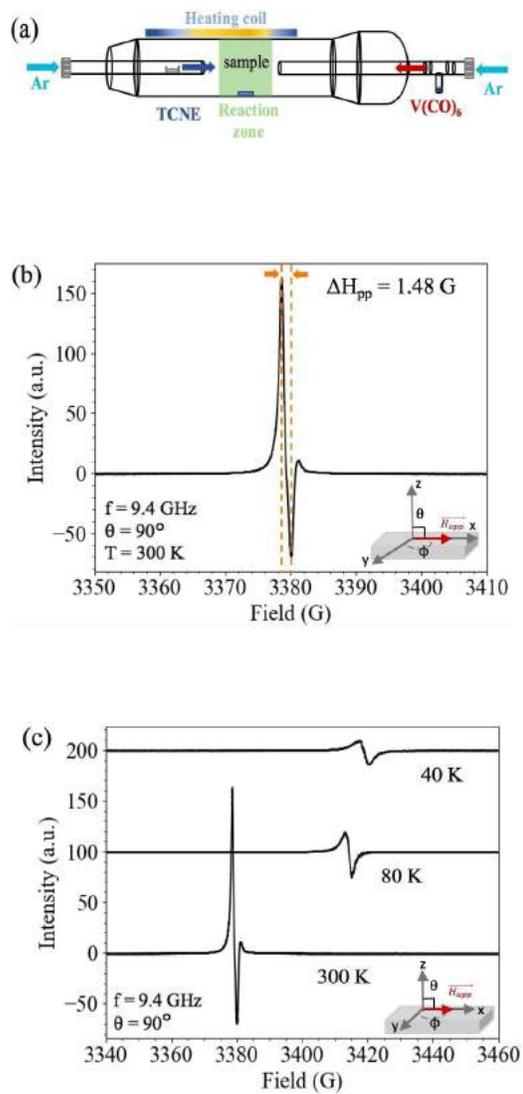

**Figure 1** H. Yusuf *et al.*

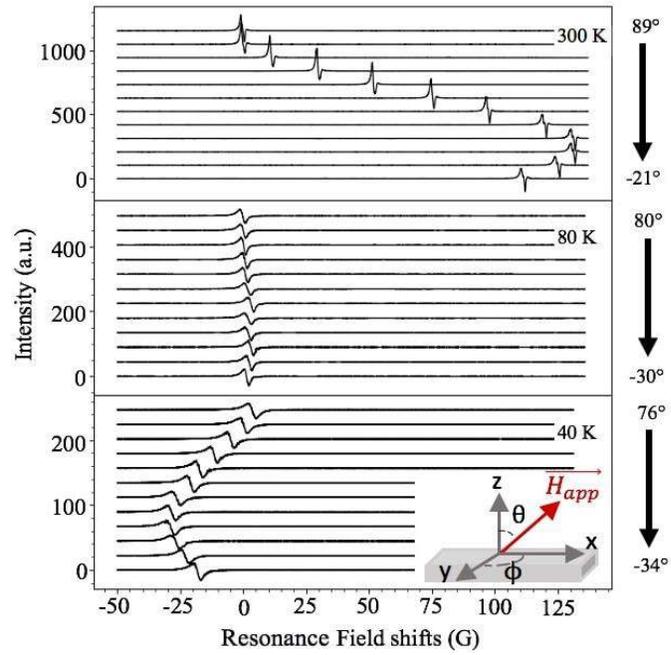

**Figure 2** H. Yusuf *et al.*

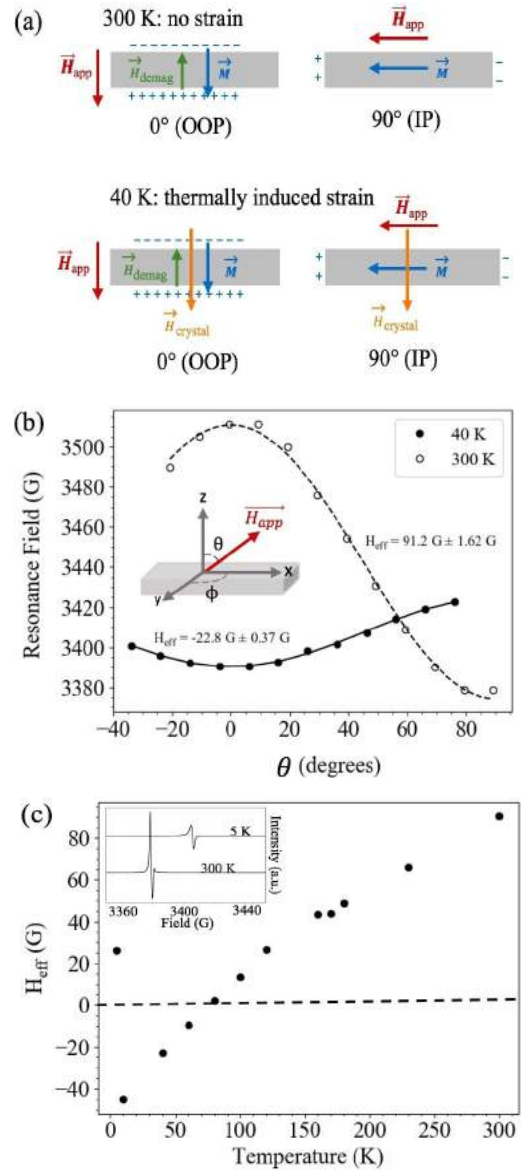

**Figure 3** H. Yusuf *et al.*

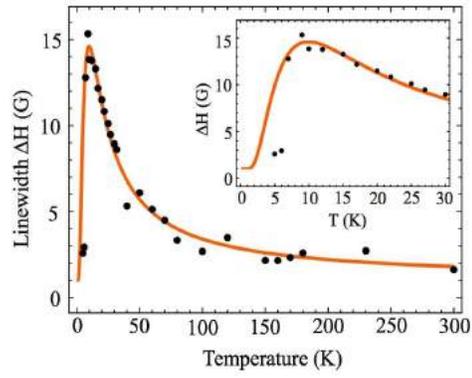

**Figure 4** H. Yusuf *et al.*

(a)

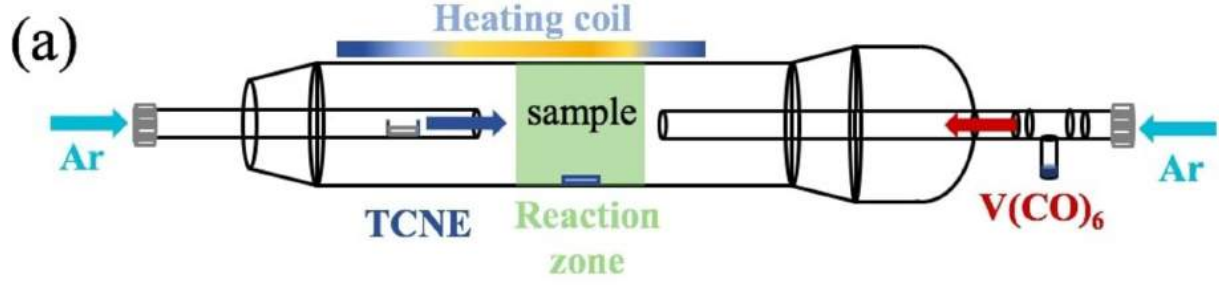

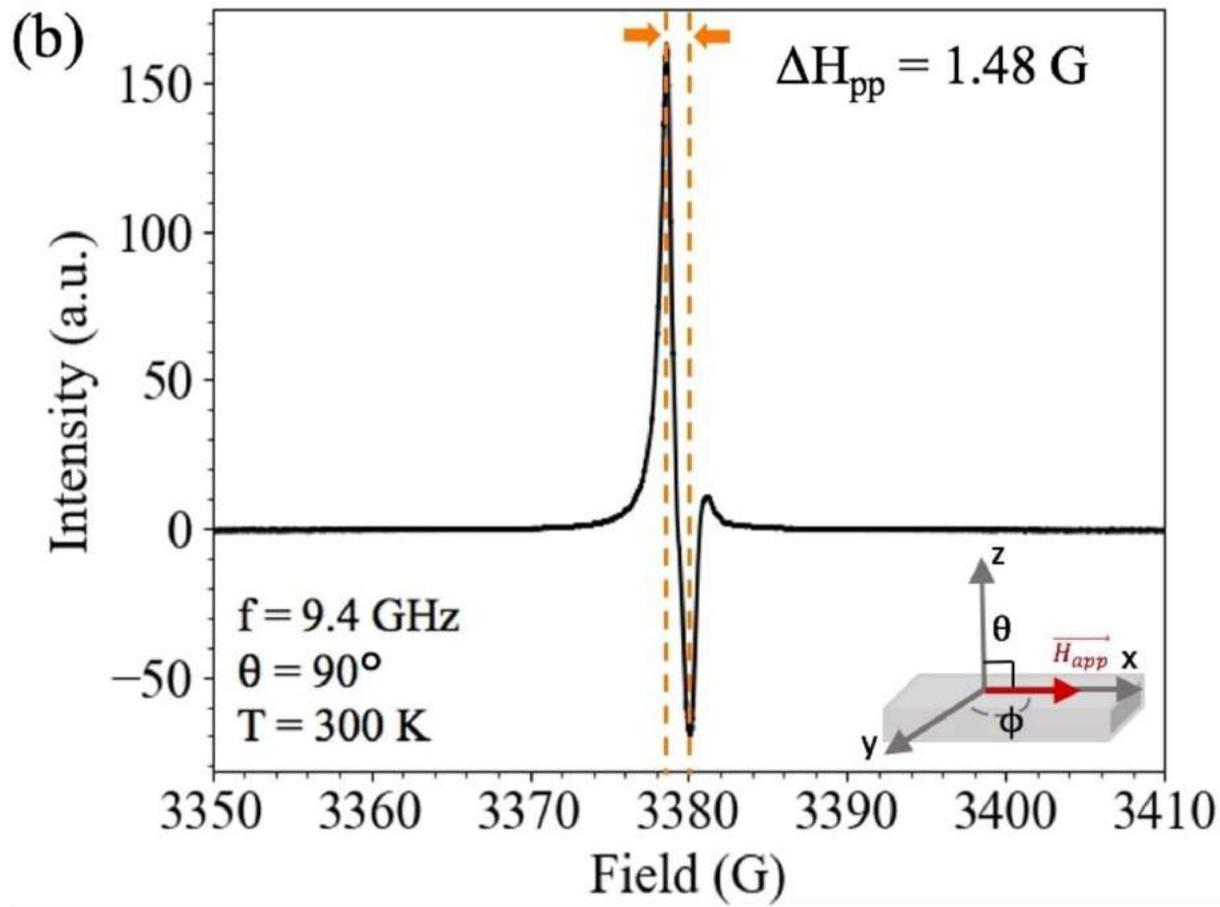

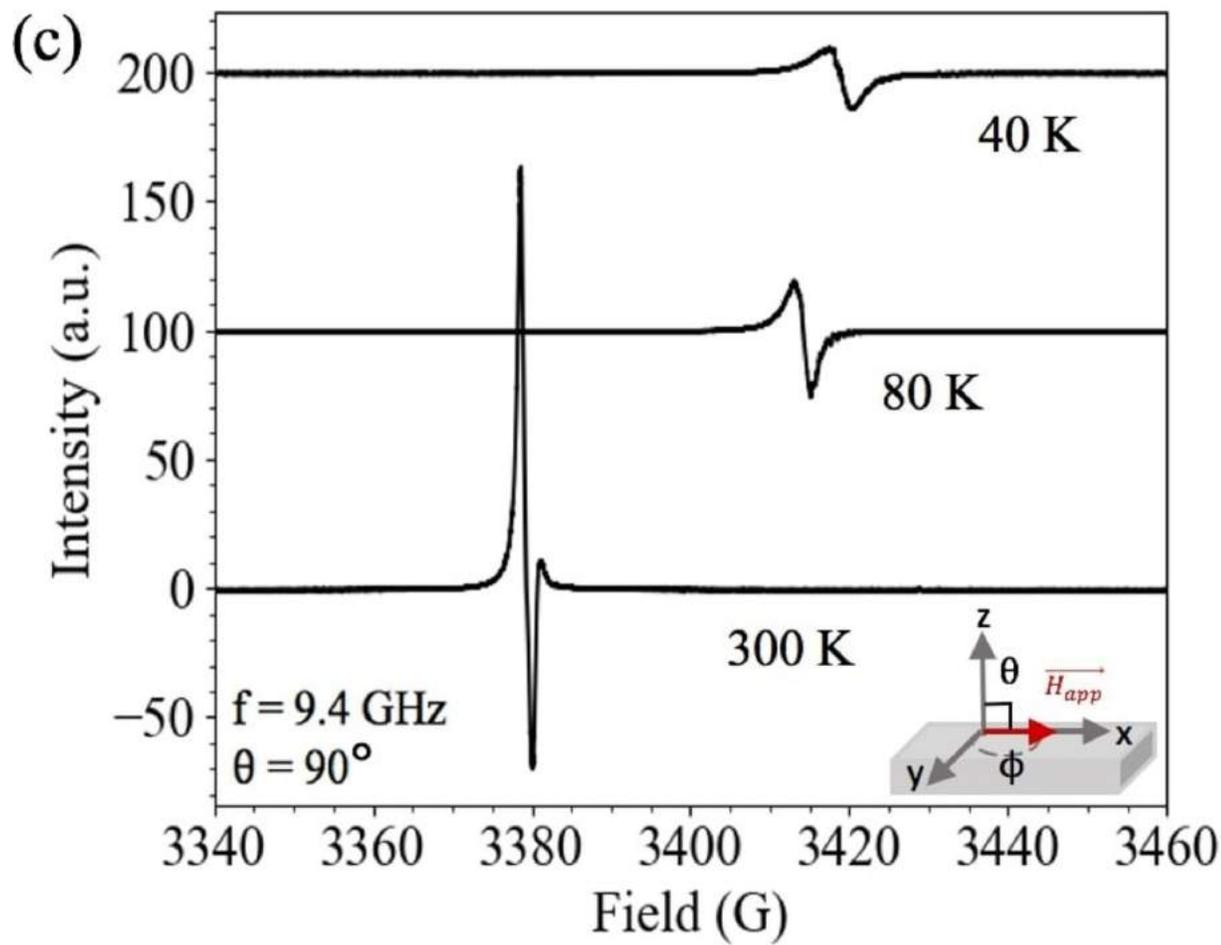

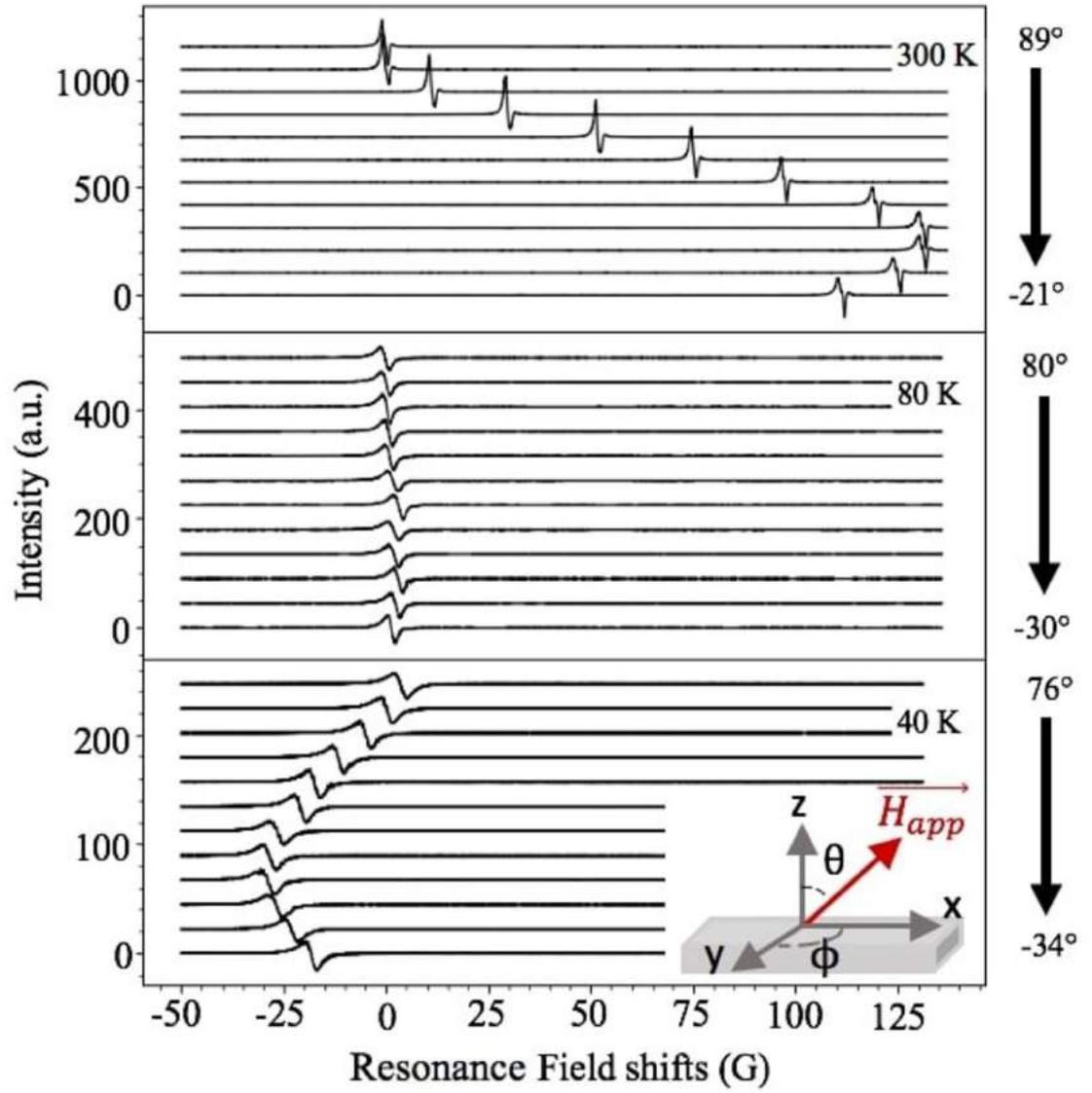

(a)

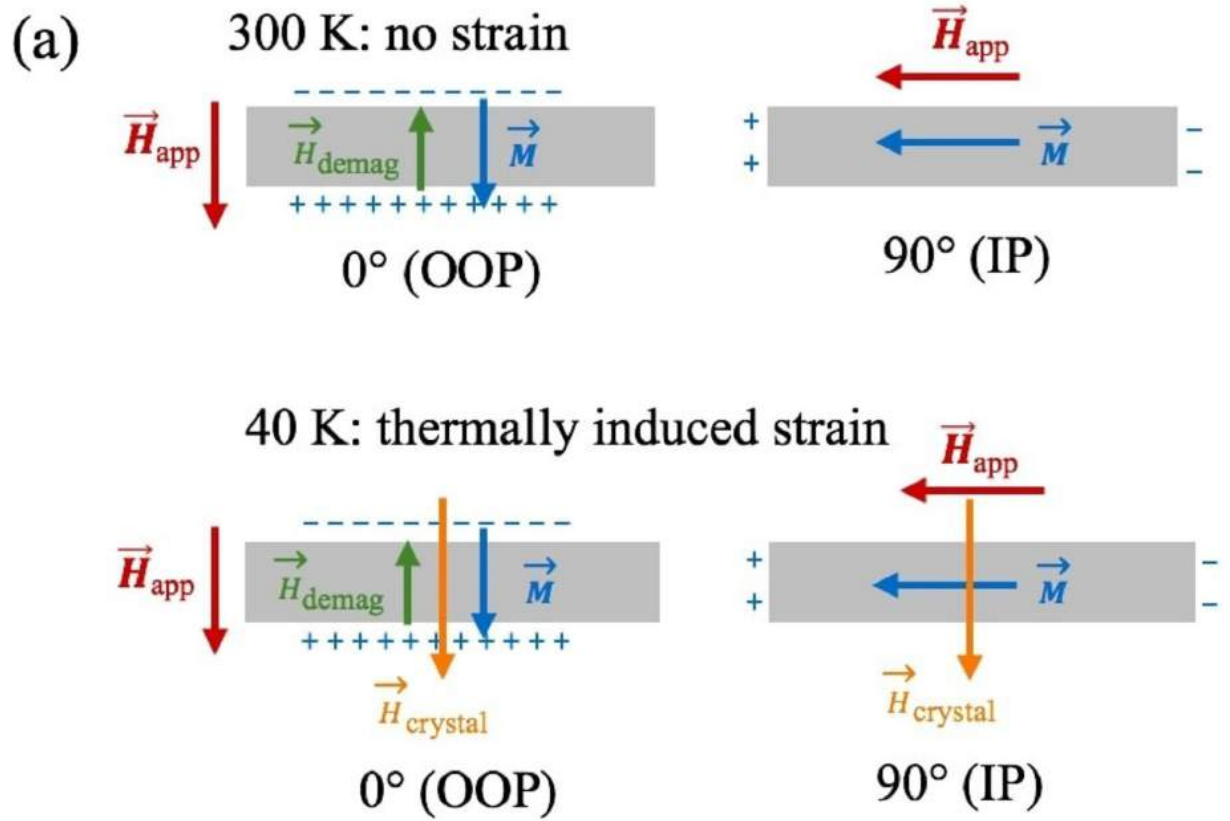

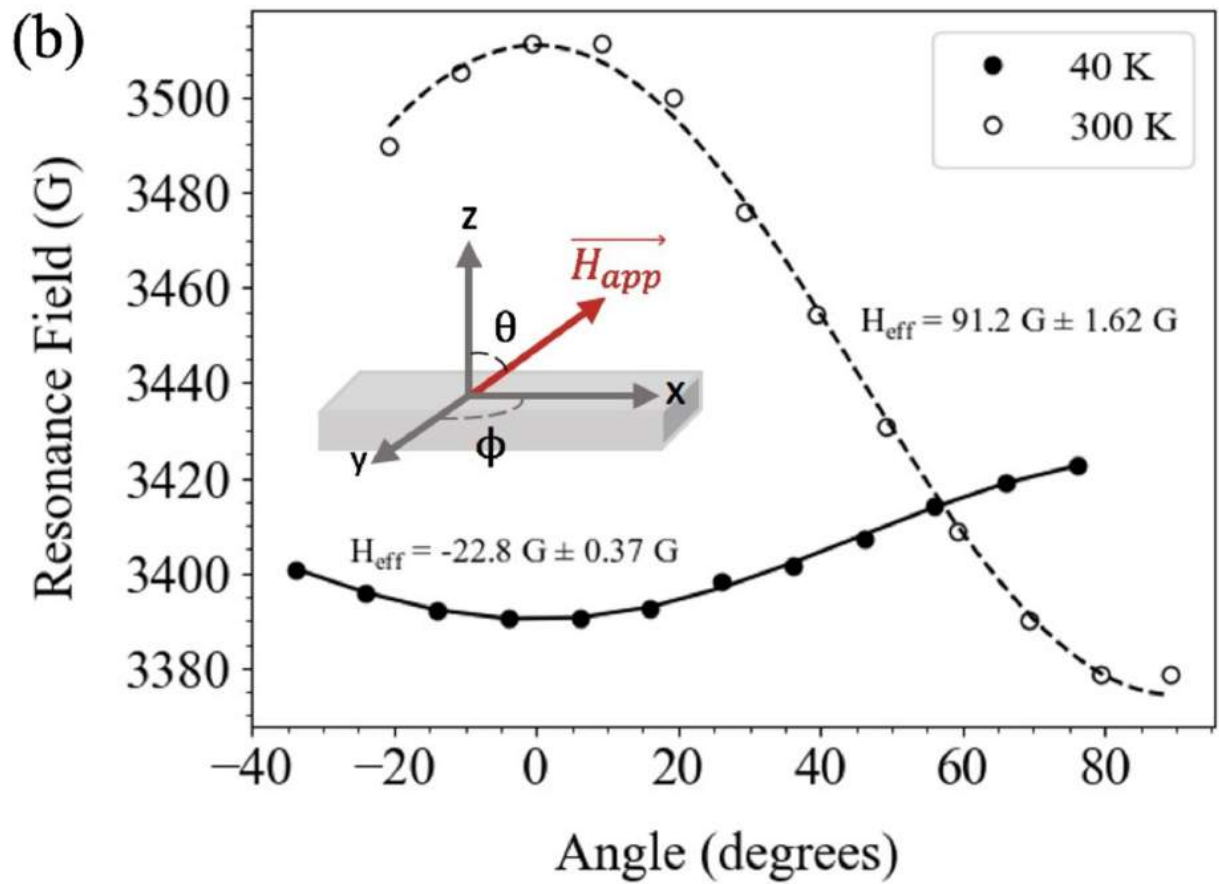

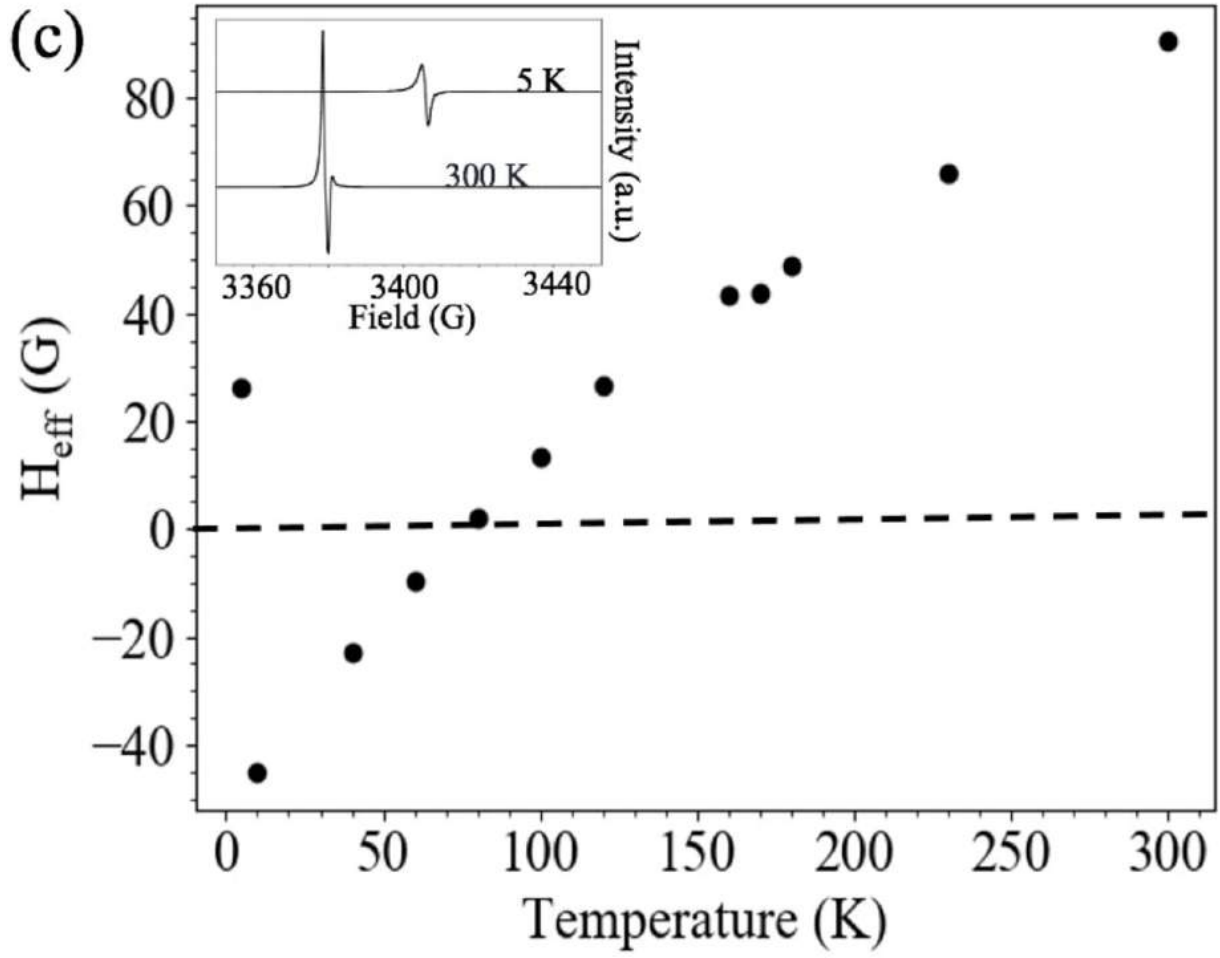

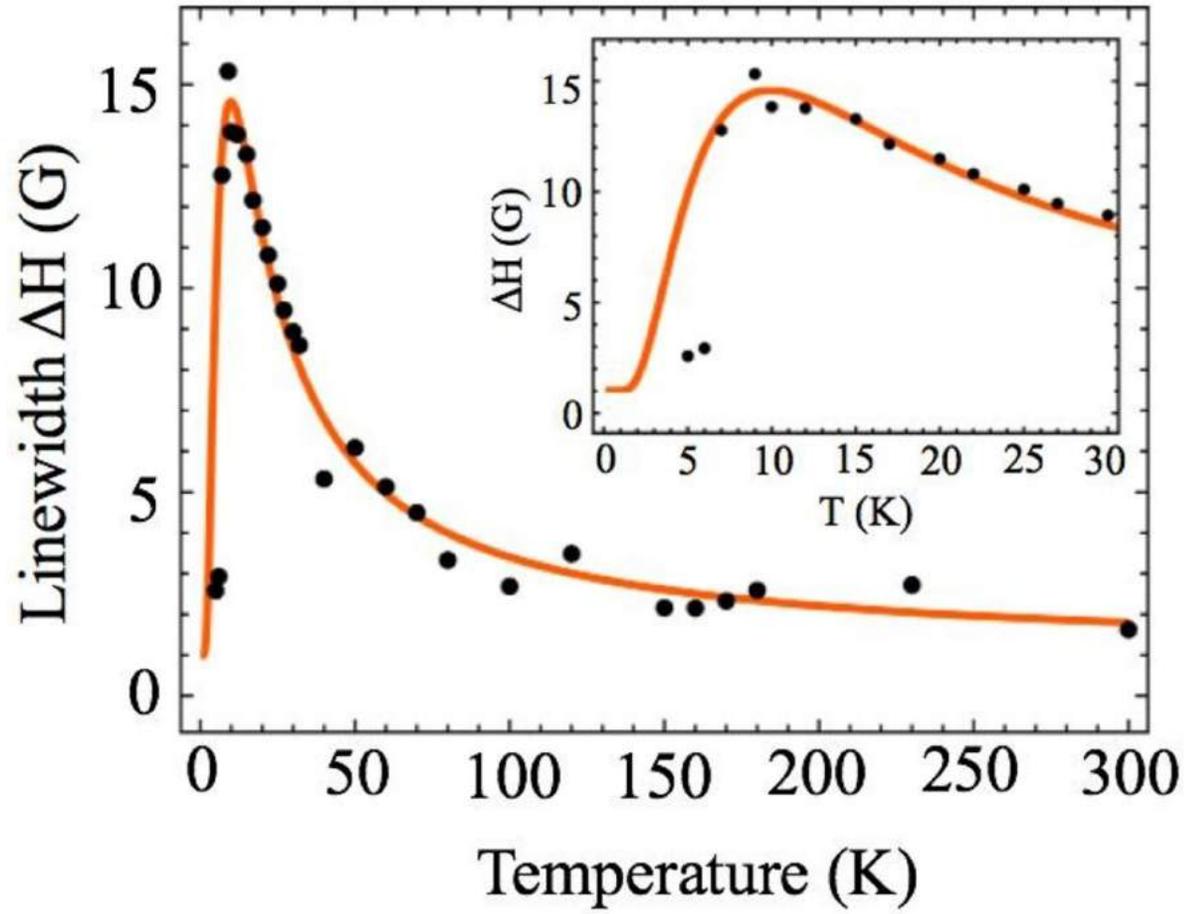

**Supplementary Materials for "Exploring a quantum-information-relevant magnonic material: ultralow damping at low temperature in the organic ferrimagnet V[TCNE]$_x$"**

H. Yusuf*[1], M. Chilcote*[1,2], D. R. Candido[3], S. Kurfman[1], D. S. Cormode[1], Y. Lu[1], M. E. Flatté[3], E. Johnston-Halperin[1]

[1]*Department of Physics, The Ohio State University, Columbus, Ohio 43210*

[2]*School of Applied and Engineering Physics, Cornell University, Ithaca, New York 14853*

[3]*Department of Physics and Astronomy, University of Iowa, Iowa City, Iowa, 52242*

* *These authors contributed equally to this work.*

## 1. Temperature-dependent anisotropy of V[MeTCEC]$_x$

Here, we investigate the magnetic properties of vanadium methyl tricyanoethylene carboxylate V[MeTCEC]$_x$ thin-films using temperature-dependent cavity ferromagnetic resonance (FMR). The MeTCEC ligand is similar to the TCNE described in the main text, and these results demonstrate that strain-dependent anisotropy is a general feature of this class of metal-ligand materials. Figure S1a shows the molecular structures of both the TCNE molecule and the MeTCEC molecule discussed below. Figure S1b shows temperature-dependent magnetization data for zero field-cooled (ZFC; open black squares) and field-cooled (FC; open red circles) measurements, and electron transport data (filled black squares) collected for V[MeTCEC]$_x$ thin-films on the same temperature axis. Notice that the maximum in the ZFC magnetization curve – sometimes referred to as the blocking temperature[13,14] – corresponds to the rapid rise observed in the resistance data. This change in electronic and magnetization properties has been associated with carrier freeze out and a magnetic phase transition in related materials such as magnatites, but in light of the results presented in the main text we note that a structural transition associated with increased strain in the films may also play a role in these measurements.

V[MeTCEC]$_x$ samples are deposited on Al$_2$O$_3$(0001) substrates using a previously reported synthesis and chemical vapor deposition (CVD) growth process.[4,16] During the deposition, argon gas carries the two precursors, MeTCEC and V(CO)$_6$, into the reaction zone where V[MeTCEC]$_x$ is deposited onto one or more substrates. The system employs three independently temperature-controlled regions for the MeTCEC, V(CO)$_6$, and reaction zone with typical setpoints of 55 °C, 10 °C, and 50 °C, respectively and with typical flow rates for each precursor of 50 sccm. Sample growth, manipulation, and handling is performed in an argon glovebox (O$_2$ < 1.0 ppm; H$_2$O < 1.0 ppm).

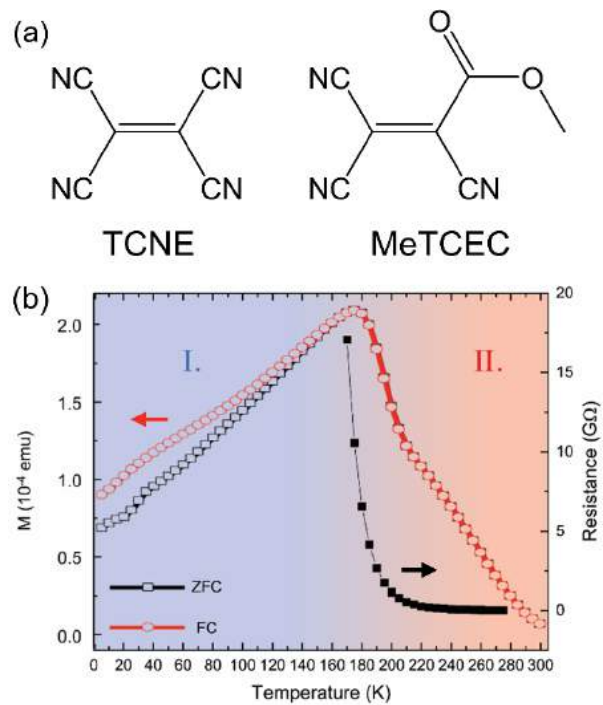

**Figure S1 (a)** The molecular structures of tetracyanoethylene (TCNE) and tricyanoethylenecarboxylate (MeTCEC). **(b)** Magnetization vs. temperature curves for zero field-cooled (ZFC; open black squares) and zero field-cooled (FC; open red circles) measurements. On the same temperature axis, resistance vs. temperature data is shown for a V[MeTCEC]x thin-film (filled black squares). The corresponding dependent-axis is shown on the right axis. Note the maximum in the magnetization data corresponds to the rapid rise observed in the resistance data.

After growth, samples are mounted onto custom microwave compatible sample holders in the appropriate orientation, protected from undesired rotation, and flame-sealed in evacuated electron paramagnetic resonance (EPR) grade quartz tubes without exposure to air. When not being measured, the sealed samples are stored in a -55 °C freezer and are found to be stable for weeks.

Figure S2 shows four ferromagnetic resonance (FMR) spectra of V[MeTCEC]$_x$ oriented both in plane (90°; see inset to Fig. 3c) and out of plane (0°) at 140 K and 80 K. The FMR response of magnetic materials is sensitive to the local field environment of the sample and therefore allows for sensitive characterization of the anisotropy fields in V[MeTCEC]$_x$. FMR measurements are performed using a Bruker electron paramagnetic resonance spectrometer setup for X-band measurements with 200 µW of applied microwave power and fitted with an Oxford Instruments ESR900 cryostat insert. The cryostat is cooled by flowing liquid nitrogen and operates at temperatures ranging from 80 K to 300 K with better than 50 mK stability during FMR measurements. In standard operation, the microwave frequency of the spectrometer is tuned between 9 and 10 GHz for optimal microwave cavity performance before the measurement, and then the frequency is fixed while the DC field is swept during the measurement.

Figure S2a shows FMR spectra collected at 140 K for the magnetic field applied in plane ($\theta = 90°$) and out of plane ($\theta = 0°$). Consistent with prior FMR measurements of organic-based magnetic materials,[6,16,17] the center field associated with the resonant feature in the in-plane spectrum is at a lower field than that of the out-of-plane spectrum, and therefore the easy magnetization axis is oriented in the plane of the film. This easy-axis orientation is the expected outcome resulting from the shape anisotropy present in thin-film samples. Figure S2b also shows FMR spectra collected with the magnetic field applied in plane ($\theta = 90°$) and out of plane ($\theta = 0°$). However, this data is collected at 80 K, further below the maximum in the V[MeTCEC]$_x$ ZFC magnetization curve than the data shown in Fig. S2a. Surprisingly, the center field of the dominant resonance feature in the in-plane spectrum is at a higher field than that of the out-of-plane spectrum. This behavior seems to indicate that the sample has an easy axis oriented out of the plane of the sample; the spectra show signs of a switch in the magnetic easy axis from in plane to out of plane as it is cooled from 140 K to 80 K.

To investigate this behavior in greater detail, angular-dependent data is collected in 10° increments as the applied field is rotated from in plane ($\theta = 90°$) to out of plane ($\theta = 0°$) of the sample. The data is shown in Figs. 3a and b for 140 K and 80 K respectively. A gray dashed line is overlaid on the data to serve as a guide to the eye. The field shifts shown in Figs. S3a and S3b are consistent with those shown in Fig. 2 above. Figure S3c shows the center fields extracted from the two-angle series, emphasizing the magnitude of the change in the anisotropy.

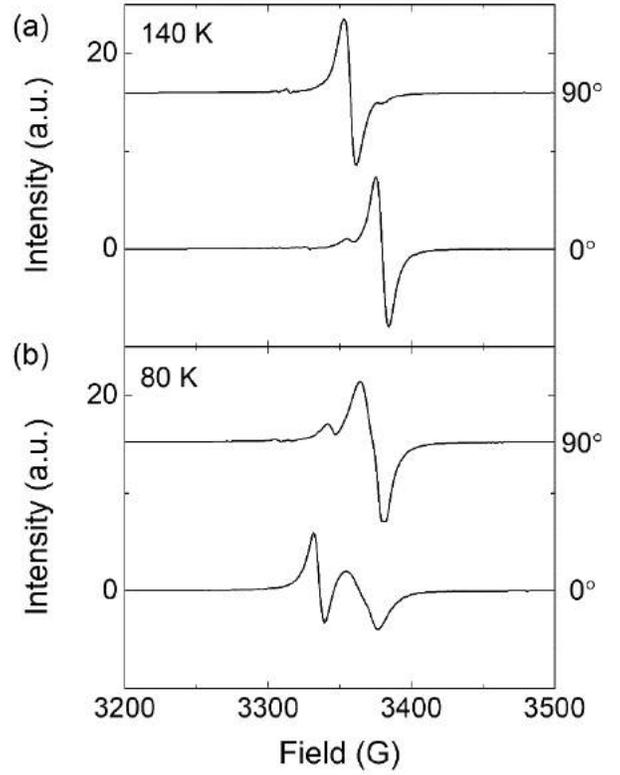

**Figure S2 (a)** Single FMR line scans at 140 K for a sample oriented in-plane (90°) and out of plane (0°) with respect to the externally applied magnetic field. **(b)** Single FMR line scans at 80 K for a sample oriented in-plane (90°) and out of plane (0°) with respect to the externally applied magnetic field.

This switch in the magnetic easy axis from in plane to out of plane present in the data suggests the presence of an additional contribution to the anisotropy beyond simply shape anisotropy. Previously, given the isotropic in-plane response of thin films at room temperature, additional contributions to the anisotropy had been excluded. However, the results here warrant the inclusion of an additional term $H_\perp$, which is responsible for inducing perpendicular anisotropy in thin films. This phenomenology is consistent with the measurements of V[TCNE]$_x$ thin films presented in the main text. Following that development, the angular dependence of the FMR response for in plane to out of plane rotation of a thin-film sample can therefore be described by,[17–19]

$$\frac{\omega}{\gamma} = \sqrt{(H - 4\pi M_{\text{eff}} \cos^2 \theta)\ (H - 4\pi M_{\text{eff}} \cos 2\theta)}$$
$$= \sqrt{(H - (4\pi M_S - H_A) \cos^2 \theta)\ (H - (4\pi M_S - H_A) \cos 2\theta)}, \qquad (1)$$

where ω is the resonance fequency, $\gamma$ is the gyromagnetic ratio, $H$ is the applied field, and $\theta$ is the polar angle of the magnetization. The FMR resonance fields are more than an order of magnitude larger than the typical saturation field for V[TCNE]$_x$, and therefore we have assumed that the magnetization is effectively parallel to the applied magnetic field (i.e. $\phi \approx \phi_H$ and $\theta \approx \theta_H$ where $\theta$, $\phi$ and $\theta_H$, $\phi_H$ are the polar and azimuthal angles of the magnetization $M$ and the applied bias field $H$, respectively). Also, note that the in-plane FMR response remains isotropic, with the $\phi$ dependence dropping out:

$$\begin{aligned}
\frac{\omega}{\gamma} &= \sqrt{H\left(H + 4\pi M_{\text{eff}}\right)} \\
&= \sqrt{H\left(H + \left(4\pi M_S - H_A\right)\right)}
\end{aligned} \qquad (\theta = 90°). \qquad (2)$$

The data in Fig. S3c are then fit according to the dispersion relation in Eq. 1. The fits are shown as solid and dashed lines in Fig. 3c. The 140 K data yields an $H_{\text{eff}} = 4\pi M_{\text{eff}} = 15.4$ Oe ± 0.1 Oe while fitting to the 80 K data result in an $H_{\text{eff}}$ value of -28.3 Oe ± 1.0 Oe. The negative value of $H_{\text{eff}}$ for the 80 K data means that $H_\perp > 4\pi M_S$ and that the film has perpendicular magnetic anisotropy. Note that the magnetic energy landscape, and therefore the angular dependence contained in Eq. 1, does not allow for the easy magnetization axis to take on an intermediate vector between in plane or out of plane for this set of anisotropy fields. This result also implies that prior measurements of the anisotropy of thin films are in fact measuring $4\pi M_{\text{eff}}$ rather than the bare $4\pi M_S$ as previously assumed.[17,20] However, as with previous studies of uniform thin films, it is challenging to disentangle this form of anisotropy from $4\pi M_S$, leading us to use the more general $H_{eff} = 4\pi M_{eff}$. Temperature-dependent FMR studies combined with careful DC magnetization measurements provide a promising avenue to decoupling the two anisotropy fields.

In comparing the data shown in Fig. S3a and b, also note that at lower temperatures, the resonance response becomes markedly multi-modal and appears to broaden. To investigate this behavior in greater detail, FMR data is collected over a range of temperatures with the applied field oriented in the plane of the sample. The data is shown in Fig. S4a. Note the clear shift of the resonant features towards higher field at lower temperatures as the in-plane orientation, which is the geometry being measured in this data set, changes from the easy magnetization axis to the hard magnetization axis.

The effective magnetization, $4\pi M_{eff}$, extracted from the data shown in Fig. S3 contains contributions from a perpendicular magnetic anisotropy energy. This $H_\perp$ does not arise from shape anisotropy in thin films and

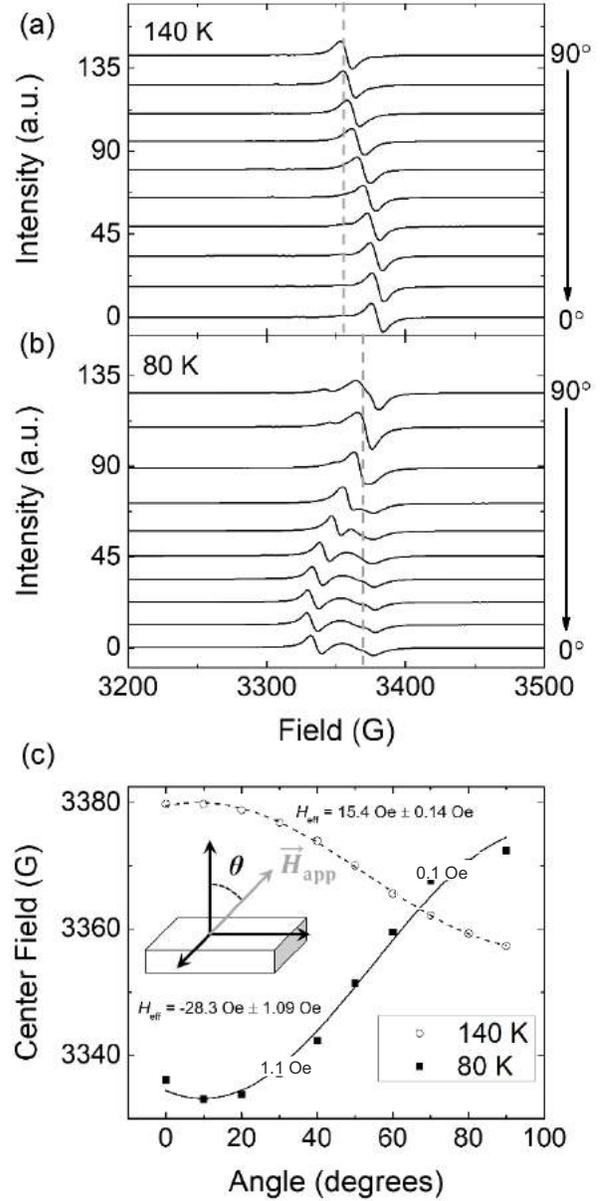

**Figure S3 (a)** Shows FMR spectra as the sample is rotated from in-plane ($\theta = 90°$) to out of plane ($\theta = 0°$) with respect to the externally applied magnetic field at 140 K. **(b)** Shows the FMR spectra as the sample is rotated from in-plane ($\theta = 90°$) to out of plane ($\theta = 0°$) with respect to the externally applied magnetic field at 80 K. **(c)** Shows the extracted center fields from the angular series shown in (a) and (b) with fits shown as solid and dashed lines. The inset shows the coordinate system with respect to the sample geometry.

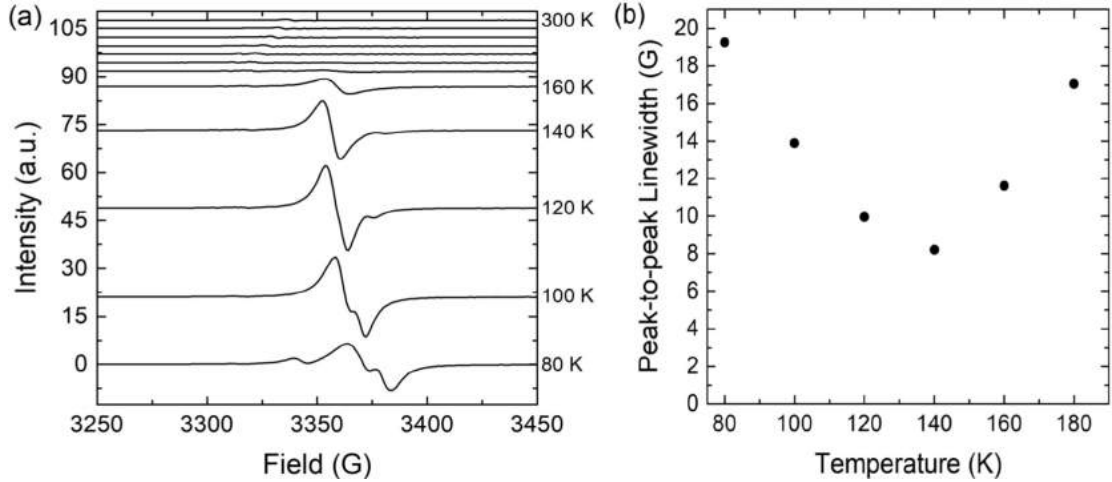

**Figure S4 (a)** Shows FMR spectra of V[MeTCEC]$_x$ sample mounted in-plane ($\theta = 90°$) with respect to the externally applied magnetic field as a function of temperature. **(b)** Shows the extracted peak-to-peak linewidths from the temperature-dependent spectra shown in (a) must instead come from a crystal-field anisotropy wherein the local exchange vector acquires some anisotropy due to some combination of lattice symmetry and strain. Given the large differences in the coefficients of thermal expansion for organic and inorganic materials (often varying by an order of magnitude or more), stain due to differential thermal expansion at the interface between the substrate and organic-based materials is likely creating an anisotropic strain field in the magnetic material. As the sample temperature is lowered, this strain field increases until $H_\perp$ becomes larger in magnitude than $4\pi M_S$ and $4\pi M_{\text{eff}}$ takes on a negative value. The result is a magnet with an easy-axis out of plane as shown in Fig. S3b.

We note that qualitatively similar results were obtained for vanadium ethyl tricyanoethylene carboxylate (V[ETCEC]$_x$). V[ETCEC]$_x$ is a third member of this class of metal ligand ferrimagnets[23,24], supporting the thesis that strain-dependent anisotropy is a common feature of this class of materials.

## 2. Method for extracting linewidth from FMR scans

The FMR scans are obtained through phase-sensitive detection, where in addition to the static DC magnetic field the sample sees a sinusoidally modulated field component that is varied at the same frequency as the amplitude modulation of the microwaves reflected from the cavity. If there is an EPR signal, that signal is converted into a sine wave whose amplitude is proportional to the derivative of the signal (change in microwave power relative to field modulation) and appears as the first derivative of a Lorentzian function. In addition, it should be noted that some FMR scans show multi peaks (for example, the 300 K scans shown in Figure 2 of the main text) and a possible reason for that could be inhomogeneous strain. As discussed in our main text, strain in our films is induced by difference in thermal expansion coefficients between V[TCNE]$_x$ and the substrate. Given that we have taken no special precautions to prevent it, we believe it is likely that this strain will be inhomogeneous, resulting in regions of our sample with differing magnetic anisotropy, and therefore the potential for additional peaks in FMR spectra. It has been reported that strain-induced distortions can alter the local electronic and crystal-field environment by changing the orbital occupancy, tilt angle between neighboring spins[25] or magnetocrystalline anisotropy[26,27], for instance, leading to local changes in magnetic anisotropy which result in the appearance of additional resonance peaks.

Since the asymmetry of the FMR lineshape and the multi-peaks need to be accounted for, scans are not simply fit by the derivative of a symmetric Lorentzian. In phase-sensitive measurements the microwave electric field generates oscillating electric currents in the sample; the oscillating magnetization due to the microwave magnetic field results in oscillating angles between the current flow and magnetization, leading to local lattice distortions which may cause the observed asymmetry in signal lineshape due to inhomogeneous broadening[28,29]. Another possible source of this asymmetry could be the result of high cavity loading[30] and the resulting phase error introduced by the automatic frequency controller of the EPR spectrometer when the sample is resonantly excited. This warrants the inclusion of a dispersion or antisymmetric term that takes into account this asymmetry, therefore the FMR scans are fitted to the sum of the derivative of an absorption (symmetric term) and dispersion (antisymmetric term) from a Lorentzian. The derivatives have the following form:

(4)

$$absorption\ derivative\ =\ \frac{-32\ \sqrt{3}\ A\ FWHM^{3}(B-B_{o})}{9\ [FWHM^{2}+4(B-B_{o})^{2}]^{2}}$$

$$dispersion\ derivative\ =\ \frac{-4\ D\ FWHM\ (B-B_{o})}{FWHM^{2}+4(B-B_{o})^{2}}$$

where $FWHM$ is the full-width at half-max, $A$ is the height of the absorption derivative, $D$ is the height of the dispersion derivative, $B_{o}$ is the location of the resonance (center) field and $B$ is the amplitude of the magnetic field that is being swept at each data point. Therefore, the resulting line shape depends on the relative contributions of these two terms.

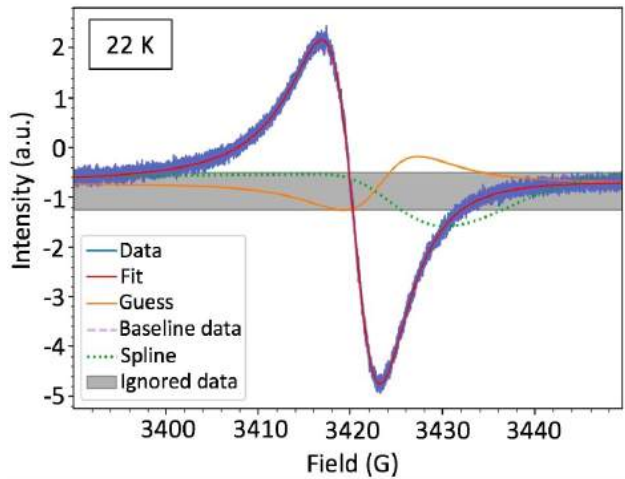

**Figure S5** Shows a single derivative fit to the FMR data collected at 22 K

For scans with an out-of-plane easy axis fitting with a single derivative sum provides good agreement with the data (Figure S5). But for scans with in-plane easy axis, due to the appearance of a modest satellite peak, obtaining a good fit to the data requires addition of up to three derivative sums. For FMR scans in the range 9 K – 80 K (out-of-plane easy axis between 9 K – 100 K and negligible anisotropy at 80 K) the date is well fit with a single derivate sum. On the other hand, fits for scans in the high temperatures between 120 K – 300 K (in-plane easy axis) give good agreement with data when two derivative sums are used, a few requiring up to three derivative sums (Figure S6b). However, FMR scans at 5 K (in-plane easy axis) and 6 K (out-of-plane easy axis) mimic the high temperature fits by requiring two derivative sums. For the purposes of this study, which explores the fundamental FMR mode, in scans showing multiple peaks we focus on the contribution from the peak that persists to low temperatures. If we plot the individual Lorentzian components of the FMR fit, we find that the first component YL1 (component with the highest overall peak-to-peak magnitude) is present in all the temperatures being considered in the range 300 – 5 K. Therefore, the linewidth date plotted against temperatures in Figure 4 of the main text is the linewidth at full width half max (FWHM) of YL1.

In Figure S6a it can be seen that fitting the FMR scan at 300 K with a single derivative sum does not provide a great fit to the data. However, from Figure S6b it becomes clear that fitting the same data with the superposition of three derivative sums or components (each with their distinct *A*, *D* and *FWHM*) gives a decent fit. In Figure S6c the amplitude of each individual component is plotted against magnetic field sweep range to provide a visual understanding of how each component contributes to the overall FMR line shape.

## 3. Temperature dependent linewidth

The V[TCNE]$_x$ linewidth dependence on temperature can be well explained from the interaction between magnons and defects or

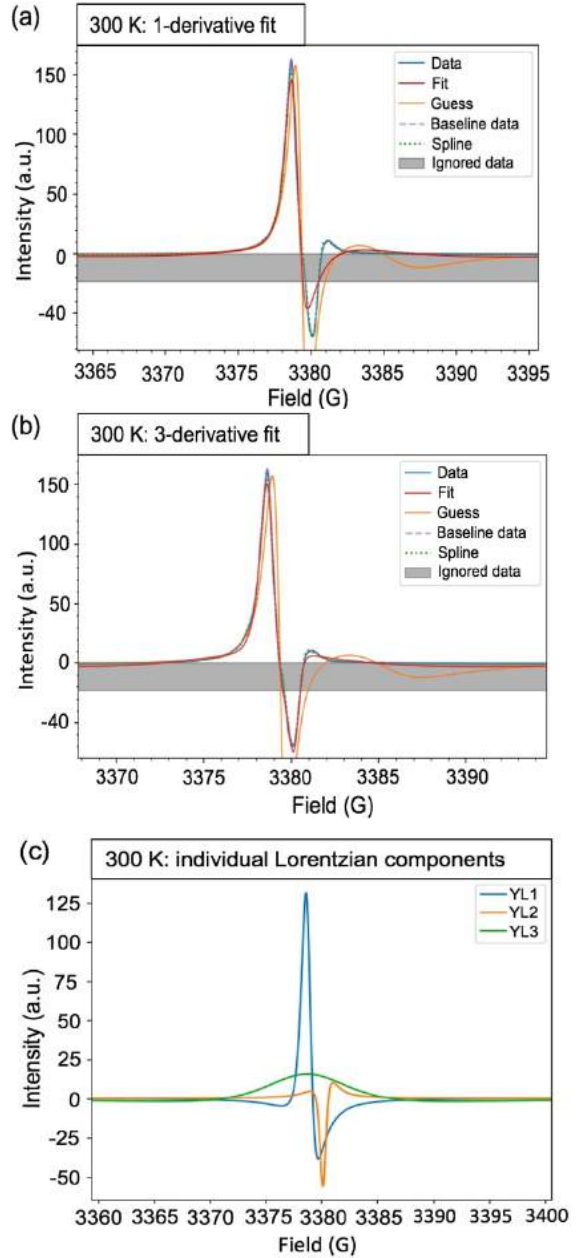

**Figure S6** (a) Shows a single derivative fit to the FMR data collected at 300 K. (b) Shows FMR scan at 300 K fitted to superposition of three Lorentzian derivative sums. (c) Amplitude of each sum or component plotted against magnetic field sweep range. YL1, YL2 and YL3 are the first, second and third components respectively.

impurities in V[TCNE]$_x$. The defects or impurities are considered to be a two-level spin systems. These experience spin-flip transitions excited by the annihilation of a uniform-magnon mode [31,32]. This process introduces a finite magnon lifetime, which in turn leads to the linewidth expression Eq. (4) in the main text. In Fig. S7, we use four different parameter sets to fit the high temperature experimental data using Eq. (4). All the different sets yield a good fitting for T>9 K, although the smaller the E$_b$, the smaller the nominal peak in

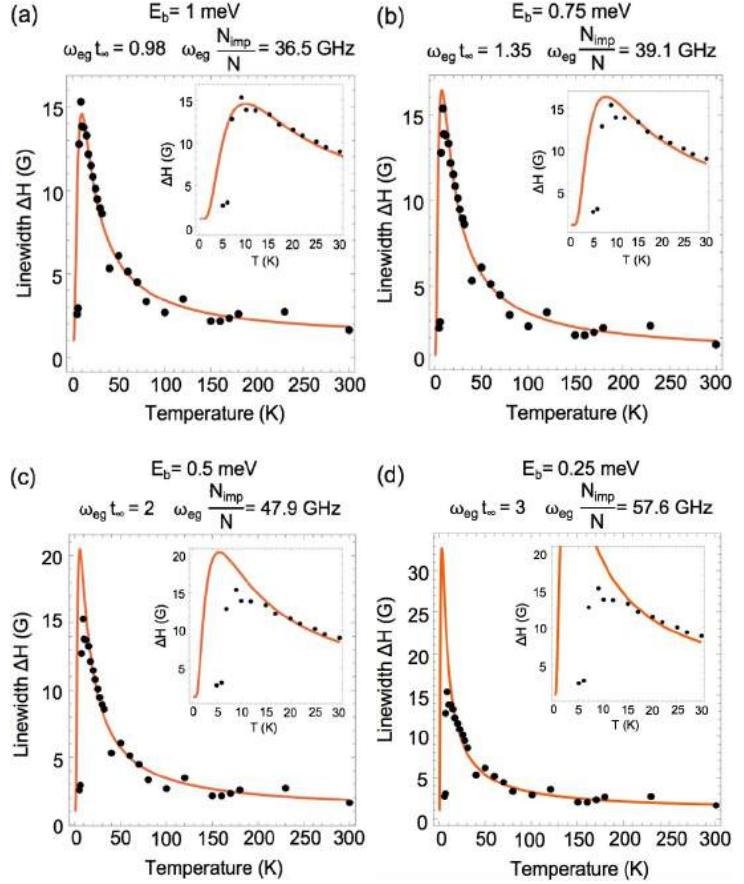

**Figure S7** (a) (b), (c) and (d) show V[TCNE]$_x$ linewidth as a function of temperature and the corresponding fit curves using fitting parameters of Eq. (4)

linewidth. As discussed in the main text, this imposes an upper bound on E$_b$ ~ 1meV.

However, it is important to note that the peak in linewidth coincides with the abrupt reversion in anisotropy from an out-of-plane easy axis to an in-plane easy axis. This change in magnetic anisotropy has the potential to have a substantial impact on spin-magnon scattering efficiency. For example, this change will result in a shift of the energy of the magnon bands (see Eq. 1 in main text), and if this change involves a commensurate change in the strain there will also be a modification to the spin-orbit coupling and exchange parameters at the paramagnetic defects. This is represented by the fits shown in Fig. S7 wherein we assume a lower temperature for the nominal peak in linewidth occurring due to spin-magnon scattering that is experimentally preempted by the change in magnetic anisotropy. As a result, we interpret the fit in Fig. 4 of the main text as an upper bound on $E_b$. The alternate fits presented in Fig. S7 agree with experimental observations at temperatures above 9 K, and therefore must be considered as possible mechanisms. Moreover, if the residual paramagnetic spins are ordered at temperatures below 9 K, one would require a large amount of energy ($>> \hbar\omega$) to populate their excited states, which is unlikely to happen. Hence, magnetic ordering of the paramagnetic spins would also enhance the suppression of spin-magnon scattering, resulting in the sharp linewidth suppression for $T < 9$ K.

**References**

[1]     C. J. Brabec, *Sol. Energy Mater. Sol. Cells* **2004**, *83*, 273.

[2]     H. Shirakawa, E. J. Louis, A. G. MacDiarmid, C. K. Chiang, A. J. Heeger, *J. Chem. Soc. Chem. Commun.* **1977**, 578.

[3]     C. W. Tang, S. A. Vanslyke, *Appl. Phys. Lett.* **1987**, *51*, 913.

[4]     Y. Lu, M. Harberts, C.-Y. Y. Kao, H. Yu, E. Johnston-Halperin, A. J. Epstein, *Adv. Mater.* **2014**, *26*, 7632.

[5]     Y. Lu, H. Yu, M. Harberts, A. J. Epstein, E. Johnston-Halperin, *J. Mater. Chem. C* **2015**, *3*, 7363.

[6]     Y. Lu, H. Yu, M. Harberts, A. J. Epstein, E. Johnston-Halperin, *RSC Adv.* **2015**, *5*, 82271.

[7]     J. L. Arthur, S. H. Lapidus, C. E. Moore, A. L. Rheingold, P. W. Stephens, J. S. Miller, *Adv. Funct. Mater.* **2012**, *22*, 1802.

[8]     J. P. Fitzgerald, B. B. Kaul, G. T. Yee, *Chem. Commun.* **2000**, 49.

[9]     J. S. Miller, A. J. Epstein, *Chem. Commun.* **1998**, 1319.

[10]    K. I. Pokhodnya, B. Lefler, J. S. Miller, *Adv. Mater.* **2007**, *19*, 3281.

[11]    E. B. Vickers, T. D. Selby, J. S. Miller, *J. Am. Chem. Soc.* **2004**, *126*, 3716.

[12]    J. Zhang, J. Ensling, V. Ksenofontov, P. Gütlich, A. J. Epstein, J. S. Miller, *Angew. Chemie Int. Ed.* **1998**, *37*, 657.

[13]   P. Granitzer, K. Rumpf, *Materials (Basel).* **2010**, *4*, 908.

[14]   R. Berger, J. C. Bissey, J. Kliava, H. Daubric, C. Estournès, *J. Magn. Magn. Mater.* **2001**, *234*, 535.

[15]   F. Cimpoesu, B. Frecus, C. I. Oprea, P. Panait, M. A. Gîrțu, *Comput. Mater. Sci.* **2014**, *91*, 320.

[16]   M. Harberts, Y. Lu, H. Yu, A. J. Epstein, E. Johnston-Halperin, *J. Vis. Exp.* **2015**, *2015*, 1.

[17]   H. Yu, M. Harberts, R. Adur, Y. Lu, P. C. Hammel, E. Johnston-Halperin, A. J. Epstein, *Appl. Phys. Lett.* **2014**, *105*, 012407.

[18]   J. Smit, H. G. Beljers., *Philips Res. Rep.* **1955**, *10*, 113.

[19]   H. Suhl, *Phys. Rev.* **1955**, *97*, 555.

[20]   N. Zhu, X. Zhang, I. H. Froning, M. E. Flatté, E. Johnston-Halperin, H. X. Tang, *Appl. Phys. Lett.* **2016**, *109*, 082402.

[21]   Y. Li, V. Coropceanu, J.-L. Brédas, *J. Phys. Chem. Lett.* **2012**, *3*, 3325.

[22]   W.-C. Wang, C.-H. Wang, J.-Y. Lin, J. Hwang, *IEEE Trans. Electron Devices* **2012**, *59*, 225.

[23]   Y. Lu, H. Yu, M. Harberts, A.J. Epstein and E. Johnston-Halperin, *RSC Adv.* **2016,** *5, 82271.*

[24]   Y. Lu, H. Yu, M. Harberts, A.J. Epstein and E. Johnston-Halperin, *J Mater. Chem. C* **2015**, *3, 7363*.

[25]   Y. Tokura, *Colossal magneto-resistive oxides, Advances in condensed matter sciences, v. 2 (Amsterdam, The Netherlands : Gordon and Beach Science Publishers,* **2000***)*.

[26]   K. Steenbeck and  R. Hiergeist,  *Appl. Phys. Lett.* **75**, *1778 (1999)*.

[27]   F. Tsui and M. C.  Smoak, *Appl. Phys. Lett.* **76**, *2421 (2000).*

[28]   Z. Celinski and B. Heinrich, *Journal of Applied Physics* **1991**, *70, 5935*.

[29]   Y. Li, F. Zeng, S.-L. Zhang, H. Shin, H. Saglam, V. Karakas, O. Ozatay, J. E. Pearson, O. G. Heinonen, Y. Wu, A. Hoffman and W. Zhang, *Phys. Rev. Lett.* **2019**, *122, 117203*.

[30]   I. B. Goldberg and H. R. Crowe, *Anal. Chem.* **1977***, 49, 9, 1353-1357.*

[31]   M. Sparks, *Ferromagnetic-Relaxation Theory (McGraw Hill, New York,* **1964***)*.

[32]   P. E. Seiden, *Phys. Rev.* **1964**, *133, A728*.